\documentstyle[prd,aps,preprint,epsf,axodraw]{revtex}
\newcommand{\gsim}{ \mathop{}_{\textstyle \sim}^{\textstyle >} }
 \newcommand{\lsim}{ \mathop{}_{\textstyle \sim}^{\textstyle <} }
\newcommand{\vev}[1]{ \left\langle {#1} \right\rangle }

\newcommand{\GEV}{~\mbox{GeV}}
\newcommand{\TEV}{~\mbox{TeV}}

\begin{document}
\tighten
\draft

 \baselineskip 0.6cm
\renewcommand{\thefootnote}{\fnsymbol{footnote}}
\setcounter{footnote}{1}
 
\title{\hfill{\normalsize\vbox{\hbox{UT-932}
}}\\
Affleck-Dine baryo/leptogenesis \\ with a 
          gauged $U(1)_{B-L}$}
 \vskip 1.2cm
\author{Masaaki~Fujii, K~Hamaguchi}
\address{Department of Physics, University of Tokyo, Tokyo 113-0033,
Japan}
\author{T. Yanagida}
\address{Department of Physics, University of Tokyo, Tokyo 113-0033,
Japan \\ and \\
Research Center for the Early Universe, University of
Tokyo, Tokyo, 113-0033, Japan}

\vskip 2cm
%
\maketitle

\vskip 2cm
\begin{abstract}
We briefly review the present status of Affleck-Dine
baryo/leptogenesis scenarios 
in the minimal supersymmetric standard model (MSSM) 
in the context of the gravity-mediated SUSY breaking,
and show that there is a 
serious cosmological problem in the Affleck-Dine mechanism.
That is, the late decay of the associated large Q-balls leads to the over
production of the lightest supersymmetric particles.
Then, we point out that the minimal extension of the MSSM by introducing
a gauged $U(1)_{B-L}$ symmetry naturally solves this problem.
Here, the breaking scale of the $U(1)_{B-L}$ can be determined
quite independently of the 
reheating temperature from the required baryon asymmetry.
It is extremely interesting that the obtained scale of the $U(1)_{B-L}$ breaking 
is well consistent with 
the one suggested from the seesaw mechanism to 
explain the recent neutrino-oscillation experiments.
We consider that the present scenario provides a new determination 
of the $U(1)_{B-L}$ breaking scale fully independent of the neutrino masses. 
We also comment on viability of the present 
scenario in anomaly-mediated
SUSY breaking models.
\end{abstract}

%

\renewcommand{\thefootnote}{\arabic{footnote}}
\setcounter{footnote}{0}

\section{Introduction} 
\label{sec:introduction}
Affleck-Dine mechanism \cite{Affleck:1985fy,Dine:1996kz} is one of the 
interesting possibilities for generating the present baryon 
(matter--antimatter) asymmetry in the early universe.
This is expected to work naturally in the supersymmetric (SUSY) 
standard model,\footnote{In most part of this paper, 
we assume the gravity-mediated SUSY breaking. 
We note on Affleck-Dine mechanism in the
anomaly-mediated SUSY breaking \cite{Randall:1999uk,Giudice:1998xp}
the last section.}since 
it has a number of flat 
directions (complex scalar fields) \cite{Gherghetta:1996dv}
carrying baryon (B) or lepton (L) charges.
A linear combination of squark and/or slepton fields, 
which we will call an AD field,
may have a large expectation value along a flat direction during an 
inflationary stage in the early universe. The AD field starts its coherent oscillation 
after the inflation ends and it creates 
a large net baryon and/or lepton asymmetry, which is 
finally transferred to matter particles when it eventually decays.

In recent developments, however, there appear several 
serious obstacles to Affleck-Dine mechanism. In particular, the 
formation of a Q-ball,
which is a kind of a 
non-topological soliton \cite{Coleman:1985ki,Lee:1992ax}, is the most serious.
The Q-balls are likely formed due to spatial instabilities of the coherent 
oscillation of the AD field \cite{Kusenko:1998si}.
It has been, in fact, shown in detailed numerical calculations that 
almost all the initial charges which the AD field carries are absorbed 
into the formed Q-balls \cite{Kasuya:2000wu,Kasuya:2000wx}. 
Thus, the resultant baryon asymmetry
must be provided by decay of the associated Q-balls, not of the AD field. 
These Q-balls have very significant consequences on Affleck-Dine
baryogenesis and cosmology \cite{Enqvist:1998si,Enqvist:1999en}.

To show how the difficulty arises from the Q-ball formation in
Affleck-Dine mechanism, let us briefly summarize the present status of
the Affleck-Dine baryo/leptogenesis scenarios in the context of the
minimal SUSY standard model (MSSM) with R-parity conservation.
The essential ingredients for Affleck-Dine mechanism to 
determine the resultant baryon asymmetry are the initial 
amplitude of the AD field (more precisely, the amplitude 
when it starts the coherent oscillation)
and the size of operators which kick this condensate to give it a 
phase rotational motion. 
An important point in this regard is that
during the inflation and in the epoch dominated by oscillations
of the inflaton, a large energy density of the universe violates 
SUSY, which induces a  SUSY-breaking mass term of
the order of the Hubble parameter $H$ \cite{Dine:1996kz} 
(we will call it a Hubble-mass term) for the AD field.
In the cases where the AD field has the minimal K\"ahler 
potential, the induced Hubble-mass term is positive,
which drives the AD field towards the origin during the inflation,
and hence the Affleck-Dine mechanism does not work. 
However, in general cases where the AD field has non-minimal couplings 
to the inflaton field, the induced Hubble-mass term can 
be negative so that the AD field develops a large expectation value.
This is crucial for Affleck-Dine mechanism to work and we assume 
this is the case in the following discussion.

The initial amplitude and evolution of the AD field 
are determined in a balance between the induced negative Hubble-mass
term and nonrenormalizable operators which lift the associated flat
direction. In the case where all of the nonrenormalizable operators in
the superpotential are forbidden by some chiral symmetries, such as
R-symmetry, the initial amplitude of the AD field is naturally expected
to be the reduced Planck scale $M_{*}\simeq 2.4\times 10^{18}\GEV$ 
which is the balance point between the induced
negative Hubble-mass term and nonrenormalizable interactions in 
the K\"ahler potential. The
operators which kick the AD field to give it a phase rotational motion 
are provided also by nonrenormalizable interactions in the K\"ahler potential.
This is the simplest and a quite possible scenario, but it generally 
predicts too much amount of baryon asymmetry. 
Furthermore, even if it is diluted by late-time 
entropy productions, the decay of the associated Q-balls 
causes a serious problem in the universe. This is because 
the decay temperature of the Q-ball,
which is inversely proportional to the
square root of its charge, is  
expected to be well below the freeze-out temperature of the lightest
SUSY particle (LSP),\footnote{See the discussion in
Section \ref{sec:The Q-balls} and the Appendix \ref{sec:appendix}.}which 
invalidates this
scenario by means of the following argument.

If the decay temperature of the associated Q-ball is lower
than the freeze-out temperature of the LSP, 
the resultant baryon asymmetry and
the abundance of the LSP cold dark matter is related by the following equation;
\begin{equation}
\Omega_{\chi}=3 \left(\frac{N_{\chi}}{3}\right)f_{B}
\left(\frac{m_{\chi}}{m_{n}}\right)\Omega_{B},\label{n-1}\end{equation}
where $N_{\chi}$ is the number of LSP's produced per one baryon number, which 
is at least $3$, and
$f_{B}\simeq 1$ is the fraction of baryon number stored in the form of 
Q-balls, and $m_{n}$ and $m_{\chi}$ are the nucleon mass and the
neutralino ($\chi$) 
LSP mass, respectively. $\Omega_{X}$ denotes the ratio of the energy
density of $X$ to the critical density of the present universe.
Using the conservative constraint on the baryon asymmetry 
from the Big-Bang nucleosynthesis, $0.004\lsim \Omega_{B}h^{2}\lsim
0.023$\cite{Olive:2000ij}, the above relation leads to a stringent  
constraint on the neutralino LSP mass as
\begin{equation}m_{\chi}\lsim 16 \GEV\;\left(\frac{h}{0.7}\right)^{2}
\left(\frac{\Omega_{\chi}}{0.4}\right)\left(\frac{3}{N_{\chi}}\right)
\left(\frac{1}{f_{B}}\right).\label{n-2}\end{equation}    
The direct experimental lower bound on the neutralino mass is 
$m_{\chi}>32.3 \GEV$ \cite{Hutchcroft:2000kb}, 
and it excludes the above scenario. 
This is an inevitable consequence of a variety of
Affleck-Dine baryogenesis scenarios\footnote{If we use purely leptonic
flat directions such as $LL\bar{e}$, the associated leptonic Q-balls (L-balls)
must evaporate above the electroweak scale to make the ``sphalerons''
effectively work, which results in a more stringent constraint.
See the discussion in Section \ref{sec:The Q-balls}.}with the
formation of large Q-balls.
Note that the Q-balls are formed when $H\sim{\cal{O}}({\rm{GeV}})$,\footnote{
See the discussion in Section \ref{sec:The Q-balls}.} which
is often the epoch dominated by the oscillating inflaton,
and the entropy productions which occur after this epoch can not change the 
size of the Q-balls, and hence their decay temperature.

On the other hand, if the general nonrenormalizable 
superpotential which is consistent with the MSSM gauge
symmetries and R-parity is present, most of the flat directions are
lifted by dimension six operators which are F-term
contributions from the dimension four
superpotential \cite{Gherghetta:1996dv},
and their phase rotational motions are caused by the associated A-terms.
Unfortunately, however, all of these A-terms, with only one interesting
exception using the $LH_{u}$ flat direction \cite{Murayama:1994em}, preserve $B-L$ and
the resultant $B$ and $L$ asymmetry are washed out by the ``sphaleron''
effects \cite{Kuzmin:1985mm}. Thus, the 
baryon asymmetry in the present universe can not be
explained.\footnote{It might be possible to make large Q-balls 
in order to
protect the produced baryon asymmetry 
from the ``sphaleron'' effects. 
We find that
it requires the existence of an unnaturally large cutoff scale 
$M\sim 10^{25}\GEV$.
Furthermore, the size of the Q-balls and the 
reheating temperature of inflation should be fine tuned so that one 
can avoid the afore mentioned LSP 
over-production problem.}The almost unique viable
scenario among the cases in which the initial amplitude of the AD field 
is fixed by the potential wall of the 
dimension six operators is the Affleck-Dine 
leptogenesis \cite{Murayama:1994em} using the 
$LH_{u}$ flat direction.\footnote{
In this scenario, the associated L-balls are expected to  
evaporate completely soon after their formation, 
or even not to be formed due to the strong thermal effects.
Even if the the AD field starts its coherent oscillation when
it is decoupled from the surrounding plasma, 
L-balls are not formed or very 
small because of the particular behaviors of the soft
SUSY-breaking mass of this flat direction under renormalization
group equations \cite{Enqvist:1998si,Enqvist:2000gq}.}This scenario has some particularly interesting
features. The resultant baryon asymmetry is almost 
independent of the reheating 
temperature of inflation \cite{Asaka:2000nb,Fujii:2001zr} and it is
determined by only one parameter, {\it{i.e.}}
the lightest neutrino mass, which results in an important prediction on
the rate of the neutrinoless double beta decay \cite{Fujii:2001zr}.

Only one remaining possibility in the presence of general
nonrenormalizable superpotential is to use the flat directions which
can escape from the small window of the  
potential wall of the dimension six operators. 
In this case they are
lifted by dimension ten operators which are F-term contributions 
of the dimension six superpotential. Here, all of the 
associated A-terms  
violate $B-L$ and thus the present baryon asymmetry 
might be accounted for by using these flat directions. 
Unfortunately, this is not a viable scenario in the MSSM. The effective
cutoff scale governing the dimension six operators in the
superpotential is
naturally considered to be the reduced Planck scale or higher. In this case
the associated Q-balls are too large to decay above the
freeze-out temperature of the LSP \cite{McDonald:2001kg}. 
Even if we extend the MSSM into the SUSY $SU(5)$ grand unified theories (GUT),
this problem can not be avoided.
This is because all of the relevant 
flat directions are singlets under the GUT gauge group and the 
initial amplitude of the AD field (hence the size of the Q-balls),
can not be suppressed.

In this paper, we consider a possibility to extend the MSSM
gauge group.
The most likely and minimal extension of the gauge group is to introduce the
$U(1)_{B-L}$ symmetry which is supposed to be broken at a very high energy 
scale.
If this is the case, anomaly cancellation 
conditions automatically require the
existence of three right-handed neutrino chiral superfields which are
singlets under the MSSM gauge group and carry $B-L$
charge $+1$. 
One can easily imagine that these right-handed neutrinos acquire large
Majorana masses of the order of the $U(1)_{B-L}$ breaking scale.
As a big bonus, we can naturally obtain a realistic mass spectrum 
for the lighter 
neutrinos via the so-called seesaw mechanism \cite{seesaw}.
We stress, as we will see later, that this minimal and
phenomenologically desirable extension of the MSSM indeed
cures the Affleck-Dine baryo/leptogenesis scenarios.

In the following part of this paper, we perform an analysis of 
the Affleck-Dine baryo/leptogenesis scenario which 
uses
the dimension six superpotential in the context of the minimal extension of
the MSSM with a gauged $U(1)_{B-L}$ symmetry.
Surprisingly enough, all of
the relevant flat directions which can pass through the small window of 
the potential wall coming from the dimension four superpotential
have nonzero $B-L$ charges with the same sign.
Thus, the D-term contribution of the $U(1)_{B-L}$ may stop the 
AD field to run toward the Planck scale
before the dimension ten operators do so. 
This is a crucial point for our scenario.
We find that the baryon asymmetry required from the Big-Bang
nucleosynthesis can be naturally obtained with relatively low reheating
temperatures enough to avoid a cosmological gravitino problem \cite{gravitino}.
Furthermore, the initial amplitude of AD field suppressed 
by the $U(1)_{B-L}$ D-term contribution  makes the
expected Q-balls small enough to decay well above the freeze-out
temperature of the LSP. 
We also emphasize that the breaking scale of the $U(1)_{B-L}$ is almost 
uniquely determined to produce the required baryon asymmetry.
We find the breaking scale $\simeq (2-7)\times 10^{14}\GEV$.
Note that the $U(1)_{B-L}$ breaking scale is determined totally 
independently of the neutrino
masses.
It is very encouraging that the obtained
scale is quite consistent with the scale of the 
right-handed neutrino masses suggested from the seesaw mechanism to
explain the recent neutrino-oscillation experiments \cite{SK-Atm}.

\section{The scenario for baryo/leptogenesis}
\label{sec:The scenario}
In this paper, we consider a minimal extension of the MSSM in which the
gauge group is extended by introducing only $U(1)_{B-L}$, and assume this
gauge symmetry is spontaneously broken at a very high
energy scale.
We postulate that all renormalizable and nonrenormalizable 
operators allowed by gauge symmetries and R-parity exist in
the superpotential.
The relevant flat directions which can pass through the small chinks of
the potential wall of the dimension six
operators are labeled by
the following monomials of chiral superfields \cite{Gherghetta:1996dv},
\begin{equation}
\bar{u}\bar{d}\bar{d},\quad LL\bar{e},\quad 
\mbox{and a linear combination of}\;\left(\bar{u}\bar{d}\bar{d},\;LL\bar{d}\bar{d}
\bar{d}\right).\label{n-3}
\end{equation}
Here, we mean that the flat directions labeled by 
$(\bar{u}\bar{d}\bar{d},LL\bar{d}\bar{d}\bar{d})$ can pass through the
dimension six potential wall with $\vev{\bar{u}\bar{d}\bar{d}
LL\bar{d}\bar{d}\bar{d}}\neq0$,
but $\bar{u}\bar{d}\bar{d}$
and $LL\bar{e}$ can do so only when $\vev{\bar{u}\bar{d}\bar{d}LL\bar{e}}=0$.
From Eq.(\ref{n-3}), we can see that all of the flat directions carry 
nonzero $B-L$ charges with the same sign. 
Thus, the $U(1)_{B-L}$ D-term contribution from the relevant flat
directions can not be canceled out within themselves.
Consequently, if 
the vacuum expectation values of some fields
to break the $U(1)_{B-L}$ symmetry are stabilized during the inflation and
in the inflaton-dominated epoch, these flat directions 
can be lifted at the breaking scale of 
the $U(1)_{B-L}$ as shown later.
This is a crucial point for our scenario.
We also comment here that 
these flat directions are $SU(5)_{\rm{GUT}}$ singlets, and hence 
the $SU(5)_{\rm{GUT}}$ D-term contribution can not lift them.

First, let us show the evolution of the AD fields during the inflation.
Here, we see that their initial values are really of the order of 
the breaking scale of the $U(1)_{B-L}$.
In the following discussion, 
all of the superfields and their scalar components
which parameterize the relevant flat directions will be
represented symbolically by $\phi$ and just treated as a single field.
This is not accurate in the case  
where there exist multiple flat directions.
However, it does not alter
the following order-estimation of the baryon asymmetry and 
hence we use this representation for simplicity.

If the K\"ahler
potential for the associated flat directions is in the minimal 
form ({\it{i.e.}} $K=\phi^{\dag}\phi$),
the induced SUSY-breaking
mass for the AD field is positive as emphasized in Ref.\cite{Dine:1996kz},
and is given by $\delta V\simeq 3H^{2}|\phi|^{2}$.
As described in the introduction, in order to have a 
large expectation value of the AD field, non-minimal couplings for
the AD field to the inflaton are required;
\begin{equation}
K=\phi^{\dag}\phi+I^{\dag}I+
\frac{b_{\phi}}{M_{*}^{2}}\phi^{\dag}\phi I^{\dag}I
+\ldots\;,
\label{n-4}
\end{equation}
where $I$ denotes the inflaton superfield.
The induced SUSY-breaking term from the couplings in Eq.(\ref{n-4})
is approximately given by
\begin{equation}
\delta V\simeq 3(1-b_{\phi})H^{2}|\phi|^{2},
\label{n-5}
\end{equation}
and we assume $3(1-b_{\phi})\sim -1$ for simplicity.
In contrast to the ordinary pictures described in the introduction, 
in our scenario, the initial amplitude of the AD field is not
determined in the balance between the negative Hubble-mass term and 
nonrenormalizable operators.
To demonstrate our point, let us consider, for example, 
the following superpotential;
\begin{equation}
W=\lambda X(S\bar{S}-v^{2}),
\label{n-6}
\end{equation}
where $\lambda={\cal{O}}(1)$ is a coupling constant, $v$ is the breaking scale of the 
$U(1)_{B-L}$, and $X,\;S,\;\bar{S}$ are chiral superfields which are
singlets under the MSSM gauge group and carry
$0,\;+2,\;-2\;{\rm{of}}\;B-L$ charges, respectively.
Then, the relevant scalar potential to determine the initial amplitude
of the AD field is given by
\begin{equation}
V_{D}\simeq \frac{1}{2}g^{2}\left(2|S|^{2}-2\frac{v^{4}}{|S|^{2}}
-|q||\phi|^{2}\right)^{2}-H^{2}|\phi|^{2}
+3(1-b_{S})H^{2}|S|^{2}+3(1-b_{\bar{S}})H^{2}\frac{v^{4}}{|S|^{2}},
\label{n-7}
\end{equation}
where $g={\cal{O}}(1)$ is 
the gauge coupling constant of the $U(1)_{B-L}$, $q\;(<0)$ is
the $B-L$ charge of the AD filed, and $b_{S},\;b_{\bar{S}}$ are the 
non-minimal couplings for the $S,\bar{S}$ fields to the inflaton 
corresponding to the coupling $b_{\phi}$ in Eq.(\ref{n-4}). 
In Eq.(\ref{n-7}), we have eliminated $\bar{S}$ field 
by minimizing the F-term from the superpotential in Eq.(\ref{n-6});
$|\bar{S}|\simeq v^{2}/|S|$, which can be justified as long as the
$U(1)_{B-L}$ breaking scale $v$ is much larger than the Hubble parameter 
during the inflation, $H_{\rm{inf}}$.
From Eq.(\ref{n-7}), 
one can show that the minimum for $|S|$ is given by
\begin{eqnarray}
|S|^{2}\simeq\frac{|q|}{4}|\phi|^{2}+\sqrt{v^{4}+\left(\frac{|q|}{4}
\right)^{2}|\phi|^{4}}
\label{n-8}
\end{eqnarray}
and $S$ and $\bar{S}$ fields have masses $\simeq v\gg H_{\rm{inf}}$ 
around this minimum.
Thus, the $S$ and $\bar{S}$ fields faithfully track this minimum 
throughout the history of the universe, 
and the contribution to the potential for the AD field from the first term
in Eq.(\ref{n-7}) practically vanishes\cite{Murayama:1994em}.
Then, from Eqs.(\ref{n-7}) and (\ref{n-8}), 
the relevant potential for the AD field in the initial stage is 
given by
\begin{eqnarray}
V_{D}(\phi)\simeq
\left\{
\begin{array}{ll}
\left((1-b_{S})\displaystyle{\frac{3|q|}{2}}-1\right)H^{2}|\phi|^{2}
+{\cal{O}}\left(H^{2}v^{4}/\left(\displaystyle{\frac{|q|}{4}}|\phi|^{2}\right)\right)\quad\;\;\mbox{for}\;\left(\displaystyle{\frac{|q|}{4}}|\phi|^{2}>v^{2}\right)
\\
\\
-\left(1+(b_{S}-b_{\bar{S}})\displaystyle{\frac{3|q|}{4}}\right)H^{2}|\phi|^{2}
+{\cal{O}}\left(\displaystyle{\frac{H^{2}}{v^{2}}}\left(\frac{|q|}{4}|\phi|^{2}\right)^{2}
\right)\quad\mbox{for}\;\left(\displaystyle{\frac{|q|}{4}}|\phi|^{2}<v^{2}\right).
\end{array}
\right.
\label{n-9}
\end{eqnarray}
Thus, we see from Eq.(\ref{n-9}) that 
the potential minimum of the AD
field and also those of the $S$ and $\bar{S}$ are, 
in fact, of the order of the $v$ during the inflation, 
if the coupling constants $b_{S}$ and $b_{\bar{S}}$ satisfy
the following conditions;
\begin{equation}
b_{S}\lsim -1,\;{\rm{and}}\;b_{\bar{S}}-b_{S}\lsim 4,
\label{n-10}
\end{equation}
where we have used the fact that $q\simeq-1/3$.
During the inflation, the AD field evolves exponentially to this minimum 
since the effective mass of the AD field 
is of the order of the Hubble parameter.

The evolution of the AD field after the inflation is very simple.
The AD field is frozen at this initial point 
$|\phi|_{0}\simeq v$ until
the Hubble parameter becomes smaller than the soft mass for the AD field
in the true vacuum, and then, it starts the coherent oscillation.  
This is because the initial point is determined almost independently 
of the Hubble parameter, as easily seen from Eq.(\ref{n-7}).
Note that a similar argument is always possible as long as 
the superfields to break the $U(1)_{B-L}$ have
masses much lager than the Hubble parameter,
and hence the above stabilization mechanism of the  
AD field is not restricted to the specific form of the superpotential as 
in Eq.(\ref{n-6}).

Now, let us estimate the baryon asymmetry.
The nonrenormalizable operators in the superpotential which
provide the relevant A-terms are given by
\begin{equation}
W=\frac{\lambda_{1}}{M_{*}^{3}}\left(\frac{S}{M_{*}}\right)
\left(\bar{u}\bar{d}\bar{d}\right)^{2},\quad
\frac{\lambda_{2}}{M_{*}^{3}}\left(\frac{S}{M_{*}}\right)
\left(LL\bar{e}\right)^{2},
\label{n-11}
\end{equation}
where $\lambda_{1}$ and $\lambda_{2}$ are ${\cal{O}}(1)$ coupling constants.
In the following discussion, we express these superpotential operators as
\begin{equation}
W=\frac{1}{6M^{3}}\left(\frac{S}{M_{*}}\right)\phi^{6},
\label{n-12}
\end{equation} 
for simplicity, where $M\;\left(\gsim M_{*}\right)$ is the effective cutoff scale.
Then, the scalar potential for the AD field is given by
\begin{equation}
V=m_{\phi}^{2}|\phi|^{2}
+\frac{m_{3/2}}{6M^{3}}
\left(\frac{S}{M_{*}}\right)(a_{m}\phi^{6}+{\rm{h.c.}})
+\frac{1}{M^{6}}\left(\frac{|S|}{M_{*}}\right)^{2}|\phi|^{10}+V_{D},
\label{n-13}
\end{equation}
where $|a_{m}|={\cal{O}}(1)$, and $m_{\phi}$ is 
the effective soft mass for the AD field which is
expected to be of the order of the gravitino mass $m_{3/2}$.
There might exist another A-term induced by the 
energy density of the universe dominated by the oscillating 
inflaton\cite{Dine:1996kz},
\begin{equation}
\frac{H}{6M^{3}}\left(\frac{S}{M_{*}}\right)(a_{H}\phi^{6}+{\rm{h.c}}),
\label{n-14}
\end{equation}
where $|a_{H}|={\cal{O}}(1)$. 
We can safely neglect this term since it does not 
alter the order of the final baryon asymmetry.\footnote{See the related
discussion in Ref.\cite{Asaka:2000nb}}
We require the initial amplitude of the AD field, $|\phi|_{0}$, and 
that of the $S$ field, $|S|_{0}$, satisfy the condition
\begin{equation}
|\phi|_{0}\lsim \left(m_{\phi}M^{3}\frac{M_{*}}{|S|_{0}}\right)^{1/4}.
\label{n-15}
\end{equation}
Then, the third term in Eq.(\ref{n-13}) is always smaller than the
last one $V_{D}$. Thus, the dimension ten operator does not play any role in
our scenario.
If the condition in Eq.(\ref{n-15}) is not satisfied, 
the resultant baryon or lepton number density when the Q-balls are formed 
is larger than that in the case
without a gauged $U(1)_{B-L}$ symmetry, and hence 
the LSP over-production problem becomes even 
worse.\footnote{In the cases without a gauged $U(1)_{B-L}$ symmetry,
the baryon or lepton number density is $\propto M^{3/2}$.
If the condition in Eq.(\ref{n-15}) is not satisfied, the resultant 
baryon or lepton number density when the Q-balls are formed is enhanced 
by the factor $(M_{*}/|S|_{0})^{1/2}\gg1$, compared with that in the 
case without a gauged $U(1)_{B-L}$. See also Eq.(\ref{fn-1}).}

The evolution of the AD field is described by the following equation;
\begin{equation}
\ddot{\phi}+3H\dot{\phi}+\frac{\partial V}{\partial\phi^{\dag}}=0.
\label{n-16}
\end{equation} 
The baryon or lepton number density is related to the number density  
of the AD field as
\begin{equation}
n=\beta i\left(\dot{\phi}^{\dag}\phi-\phi^{\dag}\dot{\phi}\right),
\label{n-17}
\end{equation}
where $\beta$ is the baryon or lepton 
charge carried by the AD field.
From Eqs.(\ref{n-16}) and (\ref{n-17}), the evolution of the baryon or
lepton number
density is given by the following equation;
\begin{equation}
\dot{n}+3Hn=2\beta \mbox{Im}\left(\frac{m_{3/2}}{M^{3}}
\frac{S}{M_{*}}(a_{m}\phi^{6})\right).
\label{n-18}
\end{equation}
By integrating Eq.(\ref{n-18}), we can obtain the resultant
baryon or lepton number density at time $t$ as 
\begin{equation}
\left[R^{3}n\right](t)=2\beta \frac{m_{3/2}}{M^{3}M_{*}}
\int^{t}dt\;R^{3}\mbox{Im}(a_{m}S \phi^{6}),
\label{n-19}
\end{equation}
where $R$ denotes the scale factor of the universe.
From Eq.(\ref{n-19}), one can easily see that the production of 
the baryon or lepton number effectively 
terminates as soon as the AD field $\phi$
starts its coherent oscillations around the origin, since the integrand
of the
right hand side of Eq.(\ref{n-19}) decreases as fast as $\propto t^{-4}$.
Then, at this time which is denoted by the subscript $0$, the
baryon or lepton number density is given by
\begin{equation}
n(t_{0})\simeq\frac{4\beta}{9}\frac{m_{3/2}}{H_{0}}
\frac{|a_{m}||S\phi^{6}|_{0}}
{M_{*}M^{3}}\delta_{eff},
\label{n-20}
\end{equation}
where $\delta_{eff}={\rm{sin}}({\rm{arg}}(a_{m})+
{\rm{arg}}(S)+6\;{\rm{arg}}(\phi))$ represents an
effective CP violating phase and is expected to be ${\cal{O}}(1)$ 
unless the initial phase of the AD field is fine tuned.
Then, the ratio of the baryon or 
lepton number density\footnote{The lepton 
asymmetry is partially converted to the baryon asymmetry due to the
``sphaleron'' effects and the baryon asymmetry is given by $n_{B}/s=8/23
\;|n_{L}/s|$ \cite{L-to-B}. 
In this case, the evaporation of the associated Q-balls
(L-balls) must be completed by the electroweak phase transition. This is 
possible, as we will see later, as long as $T_{R}\gsim 10^{4}\GEV$.}to 
the entropy density after the
reheating process of the inflation is given by
\begin{eqnarray}
\frac{n}{s}&\simeq& \frac{\beta}{9}|a_{m}|\delta_{eff}
\frac{T_{R}m_{3/2}}{H_{0}^{3}}\frac{|S\phi^{6}|_{0}}{M_{*}^{3}M^{3}}
\nonumber\\
&\sim& 2\times 10^{-10}
\left(\frac{T_{R}}{10^{5}\GEV}\right)
\left(\frac{1\TEV}{m_{\phi}}\right)^{2}
\left(\frac{v}{5\times 10^{14}\GEV}\right)^{7}
\left(\frac{m_{3/2}}{m_{\phi}}\right)
\left(\frac{M_{*}}{M}\right)^{3}
\left(\frac{|S\phi^{6}|_{0}}{v^{7}}\right),
\label{n-21}
\end{eqnarray} 
where $T_{R}$ is the reheating temperature of the inflation, and 
at the second line,
we have used $|a_{m}|\simeq \delta_{eff}\simeq 1,\;H_{0}\simeq m_{\phi}$. 
This ratio stays constant unless additional entropy
productions take place.
Here, we stress that the breaking scale of the $U(1)_{B-L}$ can be determined
as $v\simeq (2-7)\times 10^{14}\GEV$ almost independently of the
reheating temperature of the inflation for $10^{3}\GEV\lsim T_{R}\lsim
10^{7}\GEV$ (This parameter region for the reheating temperature $T_{R}$ 
should be taken to avoid the LSP over-production and the 
thermal effects, as we will see below.). 
We can see that the initial values of the AD and $S$ fields 
satisfy the condition in Eq.(\ref{n-15}) with 
$|\phi|_{0},\;|S|_{0}\sim v$ and $M\gsim M_{*}$.
One should note that, in the second line of Eq.(\ref{n-21}), 
we have assumed the absence of the relevant thermal
effects which cause the early oscillation of the AD field. 
If the induced thermal effects dominate the effective potential  
for the AD field and cause the early oscillation {\it{i.e.}} $H_{0}>m_{\phi}$, 
the resultant baryon/lepton
asymmetry is strongly suppressed \cite{ACE}, 
as easily seen from the first line of Eq.(\ref{n-21}). 

First, let
us investigate the conditions to avoid the early oscillation due to the
thermal mass terms.
The field which couples to the AD field through the coupling constant
$f_{i}$ gets an effective mass $f_{i}|\phi|$.
If this effective mass is smaller than the temperature $T$, the
thermal fluctuations of this field produces the thermal mass term for
$\phi$, which is given by $c_{i}^{2}f_{i}^{2}T^{2}|\phi|^{2}$, where
$c_{i}$ is a ${\cal{O}}(1)$ constant.
If this thermal mass $c_{i}f_{i}T$ exceeds the Hubble parameter in the
regime $H>m_{\phi}$, it
causes the early oscillation of the AD field.
Thus, to avoid the early oscillation, 
the following two conditions should not be satisfied 
simultaneously during the regime $H>m_{\phi}$;
\begin{eqnarray}
f_{i}|\phi|_{0}<T,\quad c_{i}f_{i}T>H.
\label{n-22}
\end{eqnarray}
When the energy density of the universe is dominated by
the oscillating inflaton, its decay produces the
dilute plasma with $T\simeq (HT_{R}^{2}M_{*})^{1/4}$\cite{KT}. 
This leads to the following sufficient condition to avoid the early
oscillation of the AD field due to the thermal mass terms;
\begin{equation}
T_{R}<\frac{f_{i}}{c_{i}^{1/2}}M_{*}
\left(\frac{|\phi_{0}|}{M_{*}}\right)^{3/2}.
\label{n-23}
\end{equation}
This condition can be easily satisfied as long as $T_{R}\lsim
10^{7}\GEV$, even if the cases where the coupling constants $f_{i}=
{\cal{O}}(10^{-5})$ are present.

Secondly, let us investigate another thermal effect which is pointed out 
in Ref.\cite{SomeIssues}.
The field, which has an effective mass 
$f_{i}|\phi|>T$, changes the trajectories of the running
coupling constants of the light fields to which it couples.
This effect produces
the following potential for the AD field;
\begin{equation}
\delta V(\phi)=
a T^{4}{\rm{log}}\left.\left(\frac{f_{i}^{2}|\phi|^{2}}{T^{2}}\right)
\right|_{f_{i}|\phi|>T},
\label{n-24}
\end{equation}
where $a$ is a constant which is given by the fourth power of 
gauge and/or Yukawa
coupling constants, and it is at most $|a|={\cal{O}}(10^{-2})$.
Following the same method developed in Ref.\cite{Fujii:2001zr},
one can easily show that 
the above thermal effect does not become relevant if the 
following condition is satisfied;
\begin{equation}
T_{R}\lsim \left(\frac{1}{|a|}\right)^{1/2}
\left(\frac{m_{\phi}}{M_{*}}\right)^{1/2}|\phi_{0}|.
\label{n-25}
\end{equation}
Hence, the early oscillation or trapping caused by the above 
thermal effect can also be
avoided as long as $T_{R}\lsim 10^{7}\GEV$.
Thus, the estimation 
in the second line of Eq.(\ref{n-21}) can be applied for 
a wide range of the reheating temperature $T_{R}\lsim 10^{7}\GEV$ 
in which we are free from the cosmological gravitino problem \cite{gravitino}.
\section{The Q-ball decay}
\label{sec:The Q-balls}
In this section, we estimate the size of the
associated Q-balls and the conditions that the Q-balls can evaporate or
decay well above the freeze-out temperature of the LSP, 
which is crucial to avoid overclosing the universe.
First, let us estimate the typical size of the associated Q-balls 
following the methods in Refs.\cite{Kasuya:2000wx,Enqvist:1999en}.
The relevant scalar potential for the AD field at 
the time of Q-ball formation is given by
\begin{equation}
V(\phi)=m_{\phi}^{2}\left(1+K {\rm{log}\left(\frac{|\phi|^{2}}{M_{G}^{2}}
\right)}\right)|\phi|^{2},
\label{n-26}
\end{equation}
where $M_{G}$ is the renormalization scale at which $m_{\phi}$ is
defined, and the $K$-term represents 
the one-loop corrections dominantly from
gaugino loops, and the value of $K$ is estimated in the range from 
$-0.01$ to $-0.1$ \cite{Enqvist:1998si,Enqvist:1999en,Enqvist:2000gq}.
The instability band for the AD field can be obtained from the equations for
the linearized fluctuations and is given by\cite{Kasuya:2000wx}
\begin{equation}
0<\frac{k^{2}}{R^{2}}<3 m_{\phi}^{2}|K|,
\label{n-27}
\end{equation}
where $k$ is the comoving momentum of the fluctuations of the AD field.
The best amplified mode is given by the center of the band,
\begin{equation}
\left(\frac{k^{2}}{R^{2}}\right)_{\rm{max}}\simeq \frac{3}{2}m_{\phi}^{2}|K|,
\label{n-28}
\end{equation}
and it corresponds exactly to the Q-ball size which is estimated
analytically using the Gaussian profile of the
Q-ball\cite{Enqvist:1999en} as
$R_{Q}\simeq \sqrt{2}/m_{\phi}|K|^{1/2}$.
For the best amplified mode, the perturbations $\delta\phi$ 
grow according to the
following equation;
\begin{equation}
\left|\frac{\delta\phi}{\phi}\right|\simeq 
\left|\frac{\delta\phi}{\phi}\right|_{0}{\rm{exp}}\left(
\int dt\frac{3}{4}m_{\phi}|K|
\right),
\label{n-29}
\end{equation}
and enter in nonlinear regimes when the Hubble parameter becomes $
H_{non}=m_{\phi}|K|/2\alpha$,
where
$\alpha={\rm{log}}\left(\left|\phi/\delta\phi\right|_{0}
\right)\approx 30$. The Q-balls are formed at this time,
and hence the typical charge of a single Q-ball can be estimated as
\begin{eqnarray}
Q&\approx& \frac{4}{3}\pi R_{Q}^{3}\times n(t_{non})
\nonumber\\
&\approx&\frac{8\sqrt{2}\pi\beta}{27}\frac{|K|^{1/2}}{\alpha^{2}}
|a_{m}|\delta_{eff}
\left(\frac{m_{3/2}}{m_{\phi}}\right)
\left(\frac{M_{*}}{m_{\phi}}\right)^{3}
\left(\frac{M_{*}}{M}\right)^{3}
\left(\frac{|S\phi^{6}|_{0}}{M_{*}^{7}}\right)
\nonumber\\
&\approx&
\frac{8\sqrt{2}\pi}{3\alpha^{2}}|K|^{1/2}B \left(
\frac{M_{*}^{2}}{m_{\phi}T_{R}}\right)
\approx 10^{16}\left(
\frac{1\TEV}{m_{\phi}}\right)\left(\frac{10^{5}\GEV}{T_{R}}\right),
\label{n-30}
\end{eqnarray}
where, the subscript $non$ denotes the time when the perturbations of the
best amplified mode become nonlinear, and $B=\displaystyle{n/s}\sim
10^{-10}$, and we have used the fact that
the Q-balls are formed before the reheating process of inflation is completed.
In the third line of Eq.(\ref{n-30}), we have assumed that there are no 
additional entropy productions and have used Eq.(\ref{n-21}) to connect the 
baryon or lepton number density when $H=H_{non}$ and the present 
baryon asymmetry.
Note that the third line of Eq.(\ref{n-30}) is also applicable for
other Affleck-Dine baryogenesis scenarios as long as there are no
additional entropy productions\cite{Enqvist:1999en}.

Now, we discuss the evaporation of the Q-balls following the methods
in Refs.\cite{Laine:1998rg,Banerjee:2000mb}. 
Although the most part of a Q-ball is decoupled from the thermal bath
outside, the outer region of the Q-ball
is thermalized since particles in the surrounding plasma can penetrate
into this region. 
The radius of the thermalized region inside the Q-ball 
is estimated as
\begin{equation}
R_{T}=\gamma R_{Q}\equiv \left[{\rm{log}}\left(\frac{f_{i}|\phi(0)|}{T}\right)
\right]^{1/2}R_{Q}.
\label{n-31}
\end{equation}
Here, we have used the Gaussian profile of the Q-ball;
$|\phi(r)|\simeq |\phi(0)|{\rm{exp}}\left(-r^{2}/R_{Q}^{2}
\right)$.
Note that 
the evaporation of the baryonic or leptonic 
charge from the outer region of the Q-ball is
suppressed by diffusion effects as emphasized 
in Ref.\cite{Banerjee:2000mb} and is described by the following equation;
\begin{equation}
\Gamma_{\rm{diff}}=\frac{dQ}{dt}\simeq -4\pi D R_{T}\mu_{Q}(T)T^{2},
\label{n-32}
\end{equation}
where $D=\xi/T$ with $\xi\simeq 4-6$, and $\mu_{Q}$ is the chemical
potential of the Q-ball and it is in the range $m_{\phi}\lsim\mu_{Q}(T)
\lsim T$. 
From Eqs.(\ref{n-32}) and the relation between the cosmic time and 
temperature, one can easily show that 
the evaporation is more efficient for lower temperatures and the 
total amount of the evaporated charge is mainly provided at $T\simeq
m_{\phi}$, since for temperatures below this value,
the evaporation is exponentially suppressed by the
Boltzmann factor ${\rm{exp}}(-m_{\phi}/T)$.
Thus, the total amount of the evaporated charge 
from a single Q-ball can be estimated as
\begin{equation}
\triangle Q\approx \frac{4\sqrt{2}\pi}{\zeta^{1/2}|K|^{1/2}}\xi \gamma
\left(\frac{M_{*}}{m_{\phi}}\right)\approx 10^{18} 
\left(\frac{0.01}{|K|}\right)^{1/2}
\left(\frac{1\TEV}{m_{\phi}}\right),
\label{n-33}
\end{equation}
where $\zeta=\pi^{2}g_{*}/90$.
Here, we have assumed $T_{R}\gsim m_{\phi}$.
If the initial charge of a Q-ball is larger than this value,
the remaining charge is emitted by the decay into light fermions.
The upper bound on the decay rate into light fermions is given by
\cite{Cohen:1986ct}
\begin{equation}
\left|\frac{dQ}{dt}\right|_{{\rm{fermion}}}\leq
\frac{\omega^{3}{\cal{A}}}{192\pi^{2}},
\label{n-34}
\end{equation}
where ${\cal{A}}$ is the area of the Q-ball, and $\omega\simeq m_{\phi}$. 
The upper bound is likely to be
saturated for Q-balls with $\phi(0)$ much larger than $m_{\phi}$,
which has been found from numerical calculations \cite{Cohen:1986ct}.
However, there might be an
enhancement factor $f_{s}$ due to the decay into lighter scalars,
which is expected to be at most ${\cal{O}}(10^{3})$.
Thus, the decay rate of a Q-ball can be written 
as\cite{Enqvist:1999en}
\begin{equation}
\frac{dQ}{dt}=f_{s}\left(\frac{dQ}{dt}\right)_{\rm{fermion}}.
\label{n-35}
\end{equation} 
By integrating Eq.(\ref{n-35}), we can obtain the decay temperature of
a Q-ball in the following form;
\begin{equation}
T_{d}\lsim 2\;\GEV\left(\frac{0.01}{|K|}\right)^{1/2}
\sqrt{f_{s}}\left(\frac{m_{\phi}}{1\TEV}\right)^{1/2}
\left(\frac{10^{20}}{Q}\right)^{1/2}.
\label{n-36}
\end{equation}
If the decay temperature of a Q-ball is well above the freeze-out
temperature of the LSP, $T_{f}\approx m_{\chi}/20$, 
the annihilation of the produced LSP's
effectively takes place. We see
from Eqs.(\ref{n-30}), (\ref{n-33}) and (\ref{n-36}) that if the reheating
temperature of inflation satisfies $T_{R}\gsim 10^{3}\GEV$,
the Q-ball charge is estimated as $Q\lsim 10^{18}$, and hence we are free from the LSP over-production problem.
From Eq.(\ref{n-21}), one can see that 
this condition can be easily satisfied in our scenario with the interesting
value of the $U(1)_{B-L}$ breaking scale $v\simeq (2-7)\times
10^{14}\GEV$.
Note that the breaking scale of the $U(1)_{B-L}$ can be determined 
almost independently of the reheating temperature
of the inflation as long as $10^{3}\GEV\lsim T_{R}\lsim 10^{7}\GEV$.

\section{Discussion and conclusions}
In this paper, we perform an analysis of Affleck-Dine
baryo/leptogenesis scenarios in the context of the minimal extension of
the MSSM in which the
$U(1)_{B-L}$ symmetry is gauged.
We find that all of the relevant flat directions can be lifted at the
breaking scale of the $U(1)_{B-L}$ by the $U(1)_{B-L}$ D-term
contribution, and the 
LSP over-production problem can easily be avoided.
The required baryon asymmetry from the Big-Bang nucleosynthesis can be
obtained in a wide range of the reheating temperature of 
inflation, $10^{3}\GEV\lsim T_{R}\lsim 10^{7}\GEV$, which 
are low enough to avoid the cosmological gravitino problem \cite{gravitino}.
Surprisingly enough, 
although the breaking scale of the $U(1)_{B-L}$ symmetry is
determined totally independently of the right-handed neutrino masses,
the obtained scale from the baryon asymmetry is quite consistent 
with the scale suggested from the
seesaw mechanism to explain the recent neutrino-oscillation experiments\cite{SK-Atm}.

We also comment here other possibilities to avoid the LSP over-production 
problem in Affleck-Dine baryo/leptogenesis scenarios which use dimension six
nonrenormalizable operators in the superpotential. 
We make the associated Q-balls
sufficiently small if we can suppress 
the baryon asymmetry in the initial stage of 
the AD field oscillation (which is the relevant epoch for the formation of 
Q-balls). We find that it is, in fact, possible if 
the effective cutoff scales are lower than the order of the
$10^{16-17}\GEV$ and the reheating temperature is in the range
$ 10^{3}\GEV\lsim T_{R}\lsim 10^{5}\GEV$.\footnote{
These parameter regions for the effective cutoff scale and the reheating 
temperature can be understood as follows.
Without a gauged $U(1)_{B-L}$, 
the evolution of the AD field before it starts the coherent oscillation is 
given by $|\phi|\simeq (HM^{3})^{1/4}$, which is determined from
the negative Hubble-mass term and the dimension ten 
operators. Then, the
baryon asymmetry can be calculated following a similar method in Section 
\ref{sec:The scenario} as
\begin{eqnarray}
\frac{n_{B}}{s}&\simeq&\frac{2\beta}{9}\delta_{eff} \frac{T_{R}m_{3/2}|a_{m}|}
{M_{*}^{2}}\left(\frac{M}{m_{\phi}}\right)^{3/2}
\nonumber\\
&\sim& 10^{-11}\left(\frac{T_{R}}{10\GEV}\right)\left(\frac{m_{3/2}}{1\TEV}
\right)\left(\frac{1\TEV}{m_{\phi}}\right)^{3/2}
\left(\frac{M}{M_{*}}\right)^{3/2},
\label{fn-1}
\end{eqnarray}
where the definitions are the same 
with those in Section \ref{sec:The scenario}. 
From Eqs.(\ref{n-30}), (\ref{n-33}) and (\ref{fn-1}), we can see $M\lsim 
10^{16-17}\GEV$ is required for the evaporation of the associated
Q-balls. On the other hand, there might appear
strong thermal effects for $M\lsim 10^{15}\GEV$, and the 
above estimation might be drastically changed.}This 
may be an interesting possibility since such a 
relatively low cutoff scale is suggested in the ${\cal{M}}$ theory 
\cite{Horava-Witten}. Other possibilities such as $f_{B}\ll1$
or to use the NMSSM with a light
singlino are investigated in Ref.\cite{McDonald:2001kg}.

Finally, we point out that our scenario can also
work in the case of anomaly-mediated SUSY breaking 
\cite{Randall:1999uk,Giudice:1998xp}. 
Note that there is a general serious problem \cite{Kawasaki:2001ye} for 
Affleck-Dine baryo/leptogenesis scenarios
in the case of anomaly-mediated SUSY breaking. 
Without a gauged $U(1)_{B-L}$ symmetry, the relevant scaler potential for the AD field 
which is lifted by n-dimensional superpotential is given by
\begin{equation}
V=(m_{\phi}^{2}-cH^{2})|\phi|^{2}-\frac{m_{3/2}}{n M^{n-3}}(a_{m}\phi^{n}+{\rm{h.c}})
+\frac{1}{M^{2(n-3)}}|\phi|^{2(n-1)},
\label{n-c-1}
\end{equation}
where the gravitino mass $m_{3/2}$ is much larger than the soft mass for 
the AD field $m_{\phi}$ in the anomaly-mediation models. 
Because of the presence of the large A-terms, there appears a global
minimum displaced from the origin, $|\phi|_{\rm{min}}\sim
(m_{3/2}M^{n-3})^{1/n-2}$, and the AD field is expected to be 
trapped in this minimum during its slow rolling regime.
In our scenario, the scalar potential is given by Eq.(\ref{n-13}), but
with much larger gravitino mass.
The filed value of the top of the 
hill in front of this global minimum is given by
\begin{equation}
|\phi|_{\rm{hill}}\sim \left(m_{\phi}M^{3}\frac{m_{\phi}}{m_{3/2}}
\left(\frac{M_{*}}{\vev{|S|}}\right)
\right)^{1/4}.
\label{n-c-2}
\end{equation}
Thus, as far as the initial amplitude of the AD field is smaller than
this value $|\phi|_{\rm{hill}}$, the Affleck-Dine mechanism can work
without any difficulty,
and in fact, this is what happens in our scenario with the gauged $U(1)_{B-L}$.

\section*{Acknowledgments}

M.F. thanks T. Watari for useful discussions.  
K.H. thanks the Japan Society for the Promotion of Science for
financial support.
This work was partially supported by ``Priority Area: Supersymmetry and
Unified Theory of Elementary Particles (\# 707)'' (T.Y.).

%
      
\appendix
\section{The Q-balls in the Affleck-Dine baryo/leptogenesis scenarios
with $W=0$}
\label{sec:appendix}
In this appendix, we will briefly investigate the features of the 
Q-balls formed in the Affleck-Dine baryo/leptogenesis scenarios 
without superpotential.
In this case, the AD field develops its field value as large as the reduced
Planck scale during the inflation and starts its coherent oscillation
when the Hubble parameter becomes smaller than its soft mass.
There are no relevant thermal effects because the masses of the 
fields which couple to the AD field are much larger than 
the temperature.
The baryon and/or lepton number density when the AD field starts its 
coherent oscillation is given by
\begin{equation}
n(t_{0})\approx\beta\delta_{eff}\left(\frac{m_{3/2}}{m_{\phi}}\right)^{2}
m_{\phi}M_{*}^{2},
\label{n-a-1}
\end{equation}
where $\delta_{eff}={\cal{O}}(1)$.
Based on a similar argument which leads to Eq.(\ref{n-30}),
the typical size of a single Q-ball is estimated as
\begin{eqnarray}
Q\approx\frac{4}{3}\pi R_{Q}^{3}\left(\frac{H_{non}}{m_{\phi}}\right)^{N}
\times n(t_{0}),
\label{n-a-2}
\end{eqnarray}       
where $N=3/2$ for $T_{R}>\zeta^{-1/4}\sqrt{m_{\phi}M_{*}}$ and $N=2$ for
$T_{R}<\zeta^{-1/4}\sqrt{m_{\phi}M_{*}}$. 
Even if we assume the Q-ball is formed in the
inflaton-dominated epoch, {\it{i.e.}} $N=2$, the typical size of charge is  
$Q\sim 10^{26}$.
From Eqs.(\ref{n-33}) and (\ref{n-36}), 
it is clear that these Q-balls can not 
evaporate, and its decay temperature is well below the freeze-out
temperature of the LSP.


%
%
%
%
\newcommand{\Journal}[4]{{#1} {\bf #2}, {#3} {(#4)}} 

\end{document}